\documentclass[12pt]{iopart}

\usepackage{graphicx}
\usepackage{dcolumn}
\usepackage{bm}
\usepackage{subfigure}
\usepackage{url}
\usepackage{multirow}
\usepackage{soul}
\usepackage{soul}
\usepackage{adjustbox}

\usepackage[usenames,dvipsnames]{color}

\usepackage{ulem}

\begin{document}


\title[Collective heat transport in cubic semiconductors]{Occurrence of the collective Ziman limit of heat transport  in cubic semiconductors: scattering channels and size effects}
\author{Jelena Sjakste$^{1}$}
\author{Maxime Markov$^{1}$}
\author{Raja Sen$^{1}$}
\author{Lorenzo Paulatto$^{2}$}
\author{Nathalie Vast$^1$}
\ead{jelena.sjakste@polytechnique.edu}
\address{
$^{1}$ Laboratoire des Solides Irradi\'{e}s, CEA/DRF/IRAMIS,
Ecole Polytechnique, CNRS,  Institut Polytechnique de Paris, F91128 Palaiseau c\'edex, France\\
$^{2}$ Sorbonne Universit\'e, Museum National d’Histoire Naturelle, UMR CNRS 7590, Institut de Min\'eralogie, de Physique des Mat\'eriaux et de Cosmochimie, 4 place Jussieu, F-75005 Paris, France}
\vskip 0.05cm

\begin{abstract}

In this work, we discuss the possibility of reaching the  Ziman  conditions 
for collective heat transport  in cubic  bulk semiconductors, such as Si, Ge, AlAs and AlP. 
In natural and enriched silicon and germanium, the  collective heat transport
limit is impossible to reach due to strong isotopic scattering. However, 
we show that in hyper-enriched silicon and germanium, as well as in materials with one single stable isotope like AlAs and AlP, at low temperatures, 
normal scattering plays an important role, making
the observation of the collective heat transport possible.
We further discuss the effects of sample sizes, and analyse our results
for cubic materials by comparing them to bulk bismuth, in which second sound has been detected at cryogenic temperatures. 
We find that collective heat transport in cubic semiconductors 
studied in this work is expected to occur at temperatures between 10 and 20 K.
\end{abstract}
\maketitle

\section{\label{sec:intro} Introduction}

The study of the heat transport regimes in bulk and low dimensional materials in general, and of the
phonon hydrodynamics  in particular, currently attracts a renewed interest~\cite{Ding:2022,Sendra:2022,Beardo:2021,Ghosh:2022,Ghosh:2021,Beardo:2019,Ding:2018,Machida:2020,Machida:2018,Cahill:2015,Volz:2016,Chang:2008,Yang:2010},
both from theoretical and experimental viewpoints. The discussion of recent advances can be
found in the review article of Ref.~\cite{Ghosh:2022}.

In most of the litterature until very recently,  a regime in which 
phonons behave not as independent carriers but as a collective excitation~\cite{Cepellotti:2016,Cepellotti:2017} 
which manifests itself in the form of the second sound temperature wave or of the Poiseuille flow, is referred to 
as the hydrodynamic regime. 
Indeed, the hydrodynamic behaviour is expected to occur in the limit where momentum-conserving
"normal" phonon-phonon scattering processes dominate over resistive scattering processes.
However, it was pointed out recently~\cite{Sendra:2022,Beardo:2021}
that this collective Ziman limit is not the only regime in which non-Fourier  effects 
can be observed~\cite{Beardo:2021,Beardo:2019}. As wave-like heat transport  at the nanoscale~\cite{Beardo:2021,Beardo:2019} is also referred 
to as hydrodynamic heat transport in litterature~\cite{Beardo:2019}, and to avoid confusion, in this work we 
follow the clarification of Ref.~\cite{Sendra:2022} and discuss the collective (or Ziman) limit 
of heat transport, rather than hydrodynamic regime.

The heat flow regimes in suspended graphene and graphene nanoribbons were
studied very actively~\cite{Machida:2020,Li:2019,Majee:2018,Zhang:2011d,Fugallo:2014,Lee:2015,Cepellotti:2015} since the prediction, by {\it ab initio} methods,
of the occurence of collective transport regime in graphene nanoribbons~\cite{Zhang:2011d,Fugallo:2014,Lee:2015,Cepellotti:2015}.
Recently the Poiseuille flow of phonons was experimentally observed in black phosphorus~\cite{Machida:2018}, graphite~\cite{Huang:2023,Ding:2022,Ding:2018,Huberman:2019} 
and SrTiO$_3$~\cite{Martelli:2018}. 
All of these materials have particular distinct features in their phonon dispersion facilitating "normal" momentum-conserving scattering processes necessary 
to reach the collective limit. 
The first two materials, as well as graphene, belong to the group of 2D- or layered systems in which the non-linear out-of-plane fluxural phonon mode 
plays a role of the efficient scattering channel which accumulates decaying phonons from linear acoustic branches. 
SrTiO$_3$ is an "incipient ferroelectric" with a "falling" optical polar phonon mode at the Brillouin zone (BZ) center which 
strongly decreases in frequency when the temperature is decreased but, in contrast to real ferroelectrics, 
is eventually stabilized by quantum fluctuations~\cite{Rabe:2007,Gurevich:1988}. 
Moreover, recently coherent second sound waves were observed in a dense magnon gas~\cite{Tiberkevich:2019}. 
At the same time, the collective limit of heat transport   was observed only in relatively few "common" 3D materials, 
such as Bi~\cite{Narayanamurti:1972,Mezhov:1974}, solid hellium~\cite{Ackerman:1966}, and NaF~\cite{Jackson:1970,Pohl:1976} at cryogenic temperatures.
In materials such as natural Si and Ge, the dominance of resistive processes, and in particular scattering by isotopes, prevents the occurence of the  collective limit. 
We note that the importance of the isotopic scattering in the damping of the peak
of thermal conductivity at low temperatures was demonstrated in many works~\cite{Thacher:1967,Ozhogin:1996,Zhernov:2002,Lindsay:2016,Inyushkin:2004,Inyushkin:2018,Ding:2022}. 
At the same time, we note that in contrast to the collective limit, the "high-frequency", or "driftless" second sound was recently observed in bulk Si and Ge~\cite{Beardo:2021,Beardo:2019}
in a rapidly varying temperature field. As it was pointed out in Ref.~\cite{Beardo:2021}, in the latter case the dominance of the
normal scattering events is not necessary to observe wave-like heat transport, the slow decay of the energy flux being the key requirement instead.

Turning back to the collective limit of heat transport,
methods  based on the density functional perturbation theory and on the Boltzmann transport equation (BTE) 
were shown  to accurately predict the conditions of the occurence of the collective  limit 
in many materials~\cite{Cepellotti:2015,Markov:2018,Ghosh:2021,Ghosh:2022,Huang:2023}. 
 It is expected  to occur  in samples with sizes defined by the interplay 
between normal phonon scattering ($\tau_N$), phonon-boundary ($\tau_b$) and  resistive scattering ($\tau_R$) times~\cite{Markov:2018}, under the condition:
$\tau_N<\tau_b<\tau_R$.

In the present work we apply the approach similar to the one of Ref.~\cite{Markov:2018} or Ref.~\cite{Ghosh:2021}
to discuss the conditions of the occurence of the collective limit of heat transport in cubic semiconductors, such as silicon,
germanium, AlAs and AlP, as a function of isotope composition and at temperatures below 50~K.
We show that  in materials with single stable isotope, such as AlAs and AlP, or in
isotopically hyper-enriched  silicon,
the collective limit of heat transport can be reached  at temperatures between 10 and 20~K, where the normal scattering dominates.
 We show that the reasons for
the absence of the collective limit in natural silicon and germanium
 is  isotopic scattering, and that collective heat transport phenomena such as "drifting" second sound 
 could in principle be observed in hyper-enriched silicon and germanium samples, 
which become available today~\cite{Abrosimov:2017}.



\section{\label{sec:comp} Methodology and computational details}

\subsection{Methodology}
To identify the conditions for the occurence of the collective limit of phonon transport,
we use  the solution of the Boltzmann transport equation (BTE)
beyond the single mode approximation (SMA), which allows to explicitly take into account the
 repopulation of phonon states. Thus, we use a ratio $\kappa_{VAR}/\kappa_{SMA}$ between the lattice thermal conductivity obtained by the 
solution of  BTE iobtained with the variational approach (V-BTE)  and the one obtained in the single mode approximation (SMA-BTE),
as the criterium a of the occurence of the collective regime. 
We also compare the thermodynamic average scattering rate of "normal" (momentum conserving) phonon-phonon scattering processes, $\Gamma_{av}^{n}$ to resistive Umklapp phonon-phonon scattering rate, $\Gamma_{av}^{U}$, as well as to other resistive scattering rates due to isotopic scattering, and to boundary scattering~\cite{Markov:2018}.
The thermodynamic averages of different phonon-scattering rates are calculated as~\cite{Markov:2018}:
$\Gamma_{av}=\frac{\sum_{\nu}C_{\nu}\Gamma_{\nu}}{\sum_{\nu}C_{\nu}}$, where
where $C_{\nu}$ is the specific heat of the phonon
mode $\nu$.
We note that in this work, we consider isotropic cubic crystals, in which all transport directions are equivalent, and thus there is no need to consider  direction-dependent averages of scattering rates, as it has to be done in highly anisotropic materials~\cite{Ding:2018}. 


\begin{table}[t]
 \centering
 \begin{tabular}{c|c|c|c}
    \hline
    \hline
    & composition & $M_{av}$ & $g_s$\\
    \hline
    \hline
   HEN & Si$^{28}$-99.9995\%, Si$^{29}$-0.0005\% & 27.98 & $6.38
\cdot10^{-9}$ \\
   EN & Si$^{28}$-99.983\%, Si$^{29}$-0.014\%, Si$^{30}$-0.003\% & 27.98 & $3.31\cdot10^{-7}$\\
   NAT  & Si$^{28}$-92.23\%, Si$^{29}$-4.67\%, Si$^{30}$-3.10\% & 28.09 & $2.01\cdot10^{-4}$\\
    \hline
 \end{tabular}
 \caption{\label{tab:isoparameters_si} Silicon. Composition, 
 average mass $M_{av}$ (in atomic mass units) and the isotopic disorder parameter $g_s$ for 
hyper-enriched, HEN, (Ref.~\cite{Abrosimov:2017}), enriched, EN, and natural, NAT, silicon.}
\end{table}

\begin{figure*}[t]
 \centering
  \includegraphics[scale=0.5]{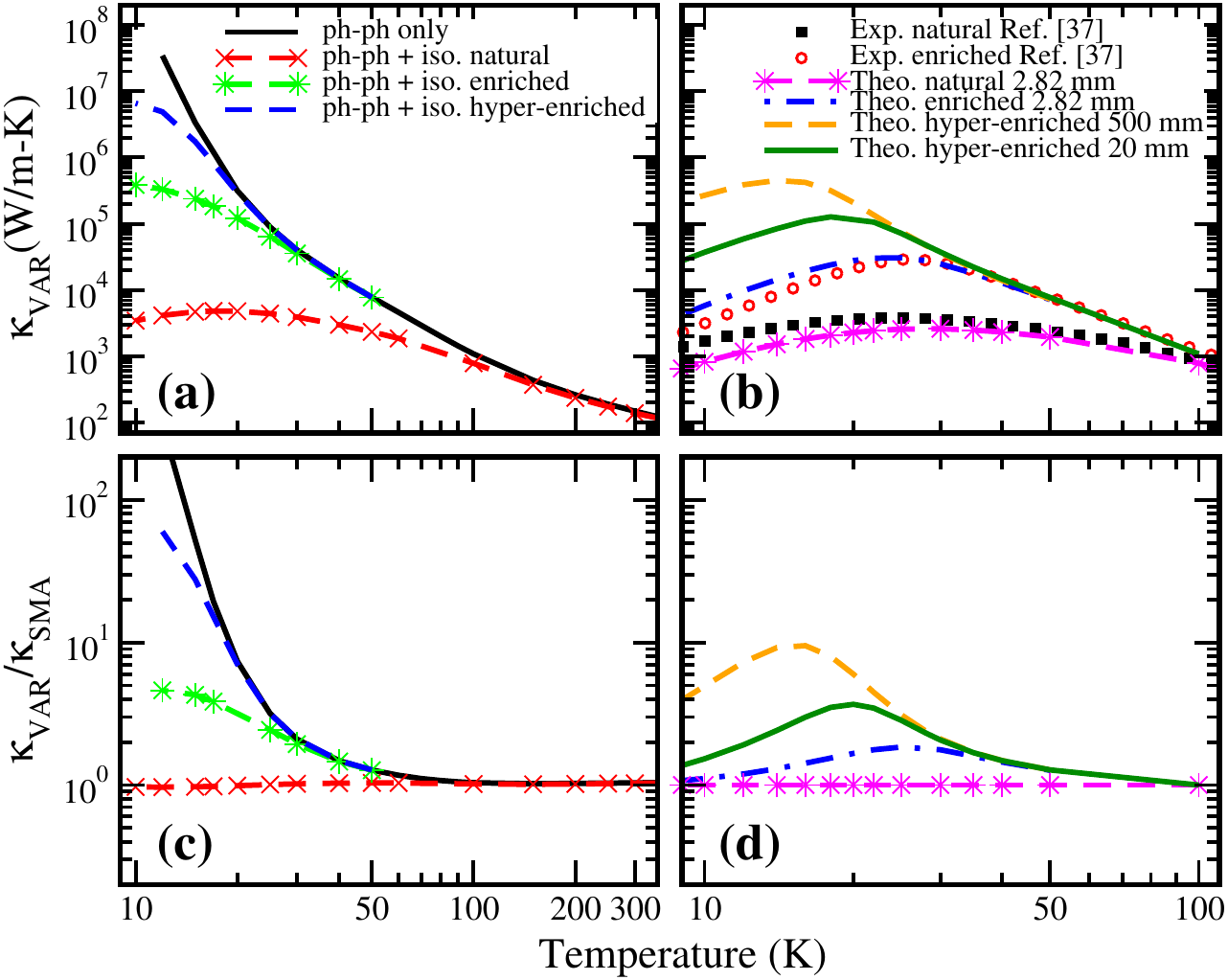}
  \caption{Panels a and c: the lattice thermal conductivity $\kappa_{VAR}$ and the ratio $\kappa_{VAR}/\kappa_{SMA}$ for Si
without boundaries.  
Black solid line - pure silicon with phonon-phonon scattering mechanism only; 
Red, blue and green dashed-dotted lines - natural, enriched and hyper-enriched silicon with phonon-phonon + isotopic scattering.  
Panels c and d:  the lattice thermal conductivity and the ratio $\kappa_{VAR}/\kappa_{SMA}$ for Si with boundaries, for different millimetric sample sizes.}
\label{kappa_ratio}
\end{figure*}


\subsection{Computational details}
Si,  AlAs and AlP are described within density functional perturbation theory in local density approximation (DFT-LDA) with norm-conserving pseudopotentials~\cite{Fhipp}. Harmonic force constants were computed on a 8$\times$8$\times$8 \textbf{q}-point grid using Density Functional Perturbation theory (DFPT)~\cite{Baroni:2001} as implemented in the  \textsc{Quantum ESPRESSO} package~\cite{Giannozzi:2017}. Third-order anharmonic constants of the normal and Umklapp phonon interactions
have been computed on a 4$\times$4$\times$4 \textbf{q}-point grid sampling the Brillouin zone (BZ) using the DFPT formalism as implemented in the D3Q package~\cite{Paulatto:2013}
and then  Fourier-interpolated on the denser
30$\times$30$\times$30  grid necessary for converged integrations of the phonon-phonon scattering rates~\cite{Paulatto:2013,Fugallo:2013}.
The lattice thermal conductivity has been computed with the linearized Boltzmann transport equation and the variational method (V-BTE) on a 30$\times$30$\times$30 \textbf{q}-point grid  with the smearing parameter $\sigma = 2$~cm$^{-1}$~\cite{Omini:1995,Fugallo:2013}. Phonon-boundary scattering is modeled by the Casimir model with the cylindrical geometry for millimeter-sized wires~\cite{Sparavigna:2002,Fugallo:2013,Markov:2016}, in the completely diffusive 
limit with no specularity~\cite{Markov:2018,Sen:2023}.
Isotope scattering is described with the  widely used Tamura model~\cite{Tamura:1983,Omini:1997,Sparavigna:2002,Fugallo:2013} with the scattering probability:

$$ P^{iso}_{j\mathbf{q},j'\mathbf{q'}} = \frac{\pi}{2N_{0}}\omega_{j\mathbf{q}}\omega_{j'\mathbf{q'}}\left[n_{j\mathbf{q}}n_{j'\mathbf{q'}}+\frac{n_{j\mathbf{q}}+n_{j'\mathbf{q'}}}{2}\right]\times$$ 
\begin{equation}
\times\delta(\hbar\omega_{\mathbf{q}j}-\hbar\omega_{\mathbf{q}'j'})\sum_{s}g_s\left|\sum_{\alpha}z^{s\alpha}_{\mathbf{q}j}*z^{s\alpha}_{\mathbf{q}'j'}\right|^2                                                                                                                                                                        \end{equation}
where $s$ run over all atoms, $\alpha$ is the Cartesian coordinate index, $j$ is the phonon branch index, $\mathbf{q}$ is the phonon wavevector, $z^{s\alpha}_{\mathbf{q}j}$ is the phonon eigenmode, $\omega_{\mathbf{q}j}$ is the phonon frequency, $g_s = \frac{(M_s - <M_s>)^2}{<M_s>}$ is the mass variance parameter. 

\section{\label{sec:results} Results: calculated thermal conductivity}

\subsection{Phonon state repopulation  in Si and isotope scattering effect}

In panel~a of figure~\ref{kappa_ratio}, we show the thermal conductivity (without boundary) calculated within
the V-BTE approach for natural silicon, isotopically
enriched silicon, and  extra-pure silicon which became available recently~\cite{Abrosimov:2017},
and which we will refer to as "hyper-enriched Si" in the present work. The isotopic compositions of Si studied in our work
are summarized in table~\ref{tab:isoparameters_si}. 
In panel~b, we show our results for the thermal conductivity of silicon in presence of boundary scattering.
In presence of boundary scattering, our calculated thermal conductivity of Si with various isotopic compositions
is found in good agreement
with available experimental data~\cite{Inyushkin:2004,Inyushkin:2018}. 
We note that our calculations of the lattice thermal conductivity
of Si,  as well as those of  AlAs and AlP which will be discussed later, 
are also in agreement with previous {\it ab initio} calculations~\cite{Ward:2009,Lindsay:2013b}
(previous theoretical data of Ref.~\cite{Lindsay:2013b} has been obtained for the 100-400~K temperature range).

\begin{figure*}[t]
  \centering
  \includegraphics[scale=0.5]{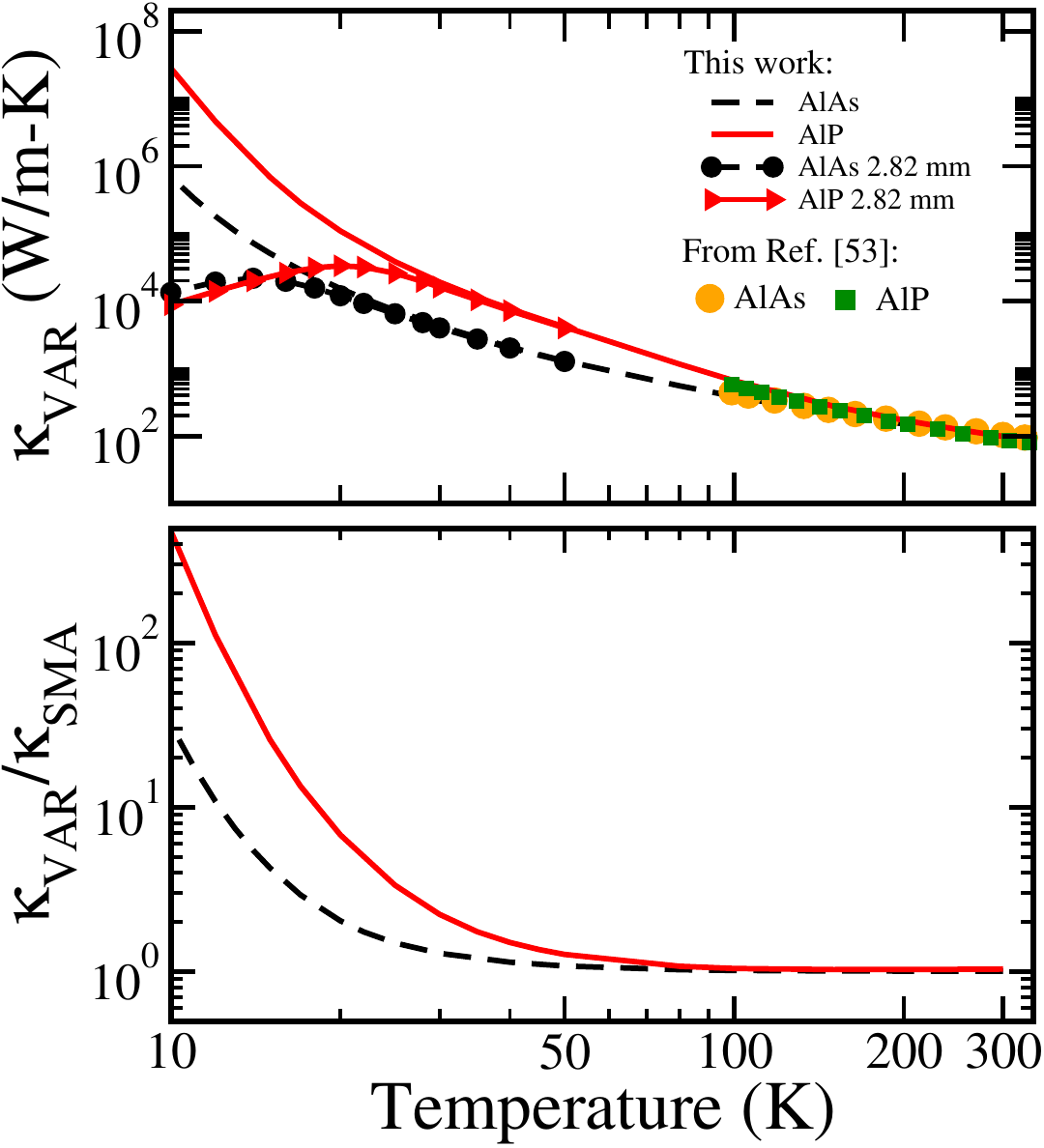}
  \caption{
Panels a and b: the lattice thermal conductivity $\kappa_{VAR}$ and the ratio $\kappa_{VAR}/\kappa_{SMA}$  for AlP and AlAs.
Black solid and dashed lines - pure AlP and AlAs respectively with phonon-phonon interaction only. 
}
\label{rates}
\end{figure*}

In panel~c  of Fig.~\ref{kappa_ratio}, 
we show the ratio $\kappa_{VAR}/\kappa_{SMA}$, for Si with various isotopic compositions.
At room temperatures and down to 70~K, the ratio $\kappa_{VAR}/\kappa_{SMA}$ is equal to one for all samples, which is a clear signature
of the kinetic regime, where heat is transported by single phonon modes.  
However, as one can see in panel~c, the ratio $\kappa_{VAR}/\kappa_{SMA}$ for isotopically pure silicon attains two orders of magnitude below 20~K (black curve), 
and similar results are obtained for hyper-enriched silicon (blue dashedcurve), demonstrating the importance
of the normal processes and of the repopulation for pure and hyper-enriched samples below 20~K.
Therefore, the possibility of the collective heat transport limit is not excluded for these samples.
 In contrast, one can see that
the collective heat transport limit is not possible in natural silicon according to our analysis (red dashed curve with crosses),  and in agreement with common knowlege.
Indeed, the $\kappa_{VAR}/\kappa_{SMA}$ ratio is close to one, and thus one can conclude that the resistive processes dominate and that the
repopulation does not play any important role in the  latter case. 
In the isotopically enriched silicon case (green dashed curve with dots),
the $\kappa_{VAR}/\kappa_{SMA}$ ratio is much larger than that of the natural silicon, however,
there is no temperature regime in which 
it  attains one order of magnitude.

 According to  our resuls,
the same conclusions are valid for germanium (results not shown): in natural and isotopically enriched samples, the
possibility of the collective heat transport limit 
which could potentially exist in hyper-enriched or pure samples,  is destroyed by isotopic scattering. 

In order to further investigate the possibility of the collective heat transport limit in the isotopically hyper-enriched silicon 
and to compare with experiments, we include boundary scattering. 
In panels~b and~d of  Fig.~\ref{kappa_ratio}, we show the thermal conductivity
of  Si and the  ratio $\kappa_{VAR}/\kappa_{SMA}$ for 
 various isotopic compositions and in presence of boundary scattering for different sample sizes.
In panel~b, our computational results for 2.82~mm sample size, for natural and enriched silicon, are 
compared with the experimental results of Ref.~\cite{Inyushkin:2004}. As one can see, the agreement
between calculated and experimental thermal conductivity is very good.
To illustrate the effect of sample size on the lattice thermal conductivity, we also show
the calculated results in hyper-enriched silicon for the 20~mm and 500~mm sizes. 

 As one can see in panel~d of Fig.~\ref{kappa_ratio}, as expected, the ratio $\kappa_{VAR}/\kappa_{SMA}$
is reduced in presence of the boundary scattering, 
for all sizes and all isotopic compositions. 
Nevertheless, normal processes  play an important role for
hyper-enriched samples, especially for sample sizes of 20~mm and larger. 

The possibility of the collective heat transport limit in isotopically enriched
samples of 2.82~mm size was discussed in Ref.~\cite{Inyushkin:2004}. 
As one can see in panel~d of Fig.~\ref{kappa_ratio}, we find 
indeed that the ratio  $\kappa_{VAR}/\kappa_{SMA}$ for those samples (blue dot-dashed curve)
exceeds 1 for temperatures between 10~K and 60~K,
with the maximum value of 1.8 at 25~K, confirming the conclusions of  Ref.~\cite{Inyushkin:2004}
that the collective heat transport can exist to some extent. However, the
effect of the repopulation of phonon states by the normal processes is strongly
reduced by the isotopic and boundary scattering for 2.82~mm samples.


\subsection{Repopulation in AlAs and AlP}

In the previous section, we have demonstrated that the isotopic stattering is the main reason why
the collective limit can not be observed in Si and Ge, while it could exist in pure or hyper-enriched samples.
 This is the reason why we have
decided to further explore cubic materials that naturally have no isotopes, such as AlP and AlAs.

In panel~a of Fig.~\ref{rates}, we show our calculated lattice thermal conductivity
for  AlP and AlAs, as a function of temperature. 
In panel~b of Fig.~\ref{rates}, in analogy with the analysis of Fig.~\ref{kappa_ratio},
we show the ratio $\kappa_{VAR}/\kappa_{SMA}$ for AlP and AlAs.
One can see that, similarly to pure Si, the $\kappa_{VAR}/\kappa_{SMA}$ ratio for AlP  attains two orders of
magnitude around 17~K. The same is true for AlAs, at slightly lower temperatures around 8~K.
Thus, we conclude that according to our results,  repopulation due to normal processes is very strong in
AlP and AlAs at low temperatures. Therefore, the collective transport limit 
can also exist in AlP and AlAs. 
Also, we can conclude that contrary to common belief, normal processes play an important role
in cubic materials such as Si, AlP and AlAs, but at temperatures below 20~K, and 
when isotopic scattering is absent or strongly reduced.
In the next section, we further analyse the effect of sample sizes.

\section{Discussion: size effects and scattering rates}

\begin{figure*}[t]
\centering
\includegraphics[scale=0.4]{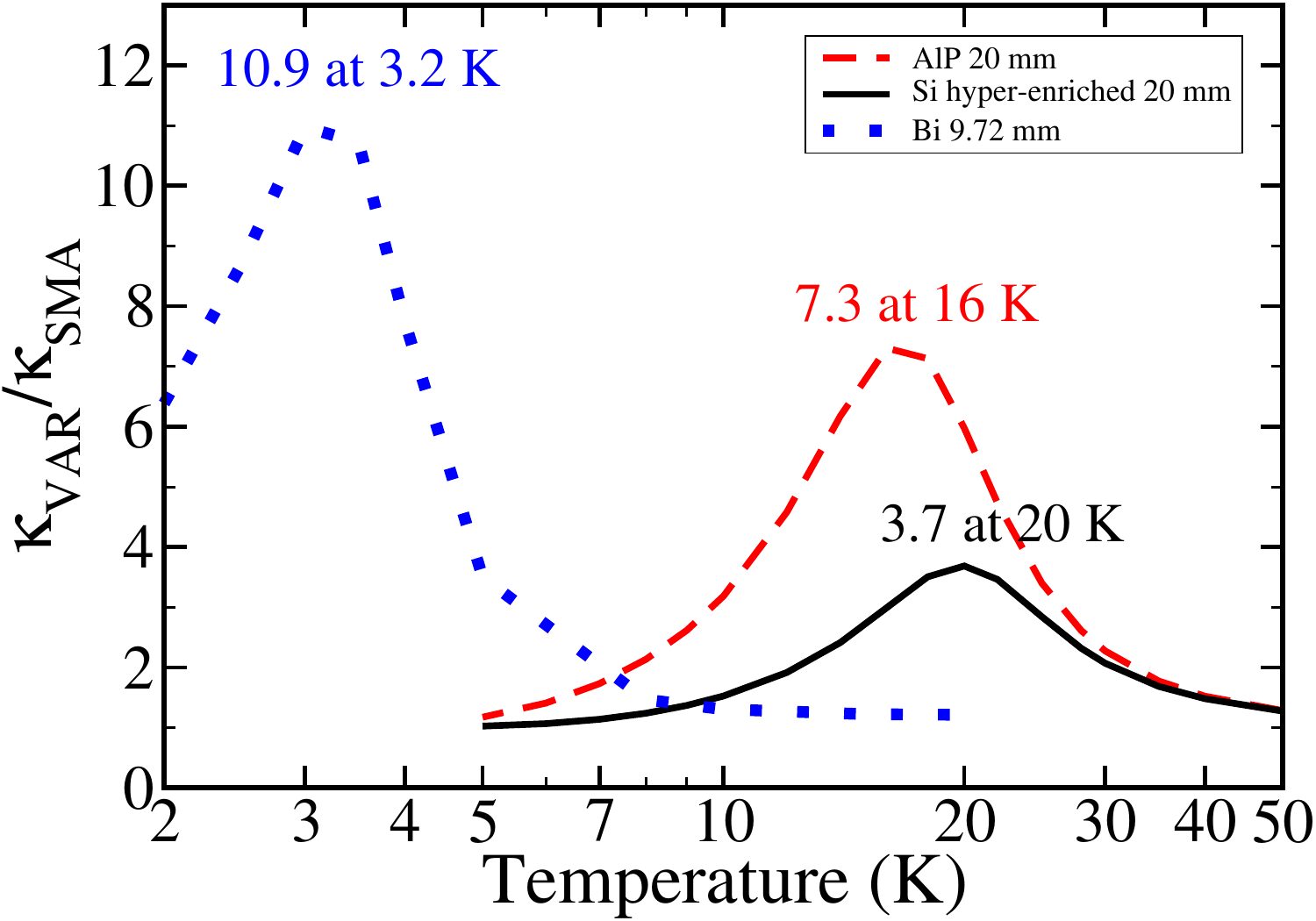}
 \caption{ $\kappa_{VAR}/\kappa_{SMA}$ ratio for AlP and hyper-enriched Si samples of 20~mm, compared
to $\kappa_{VAR}/\kappa_{SMA}$ ratio for Bi sample of 9.72~mm.}
\label{materials_size}
\end{figure*}

The effect of boundary scattering,
analysed in panel~d of Fig.~\ref{kappa_ratio} for silicon, appears
to be very strong. Indeed, the $\kappa_{VAR}/\kappa_{SMA}$ between 10 and 20~K for
hyper-enriched samples  is reduced by almost two orders of magnitude (without borders), from about one hundred 
to 3.7 in 20~mm samples (green curve).  

In Fig.~\ref{materials_size}, we compare
$\kappa_{VAR}/\kappa_{SMA}$ ratios in hyper-enriched silicon,  cubic AlP
and bismuth, which was studied in our earlier works~\cite{Markov:2018,Markov:2016}.
The interest in comparing  materials studied in the present work with Bi resides in the fact that the collective 
heat transport (drifting second sound) was experimentally observed in the latter~\cite{Narayanamurti:1972}.
As one can see in Fig.~\ref{materials_size},  scattering by boundaries reduces the repopulation effects in all
three materials. We can also note that the effect of repopulation reduction by  boundary scattering appears 
to be somewhat weaker in Bi, resulting is larger peak value of $\kappa_{VAR}/\kappa_{SMA}$ ratio.
At the same time, the peak value of $\kappa_{VAR}/\kappa_{SMA}$ in AlP at 16 K for 20~mm sample is still comparable 
to that of Bi at 3.2 K for 9.72~mm sample, indicating that observation of the
collective heat transport limit in AlP must be possible at temperatures around 16 K for 20~mm samples.

\begin{figure*}[t]
\centering
\includegraphics[scale=0.6]{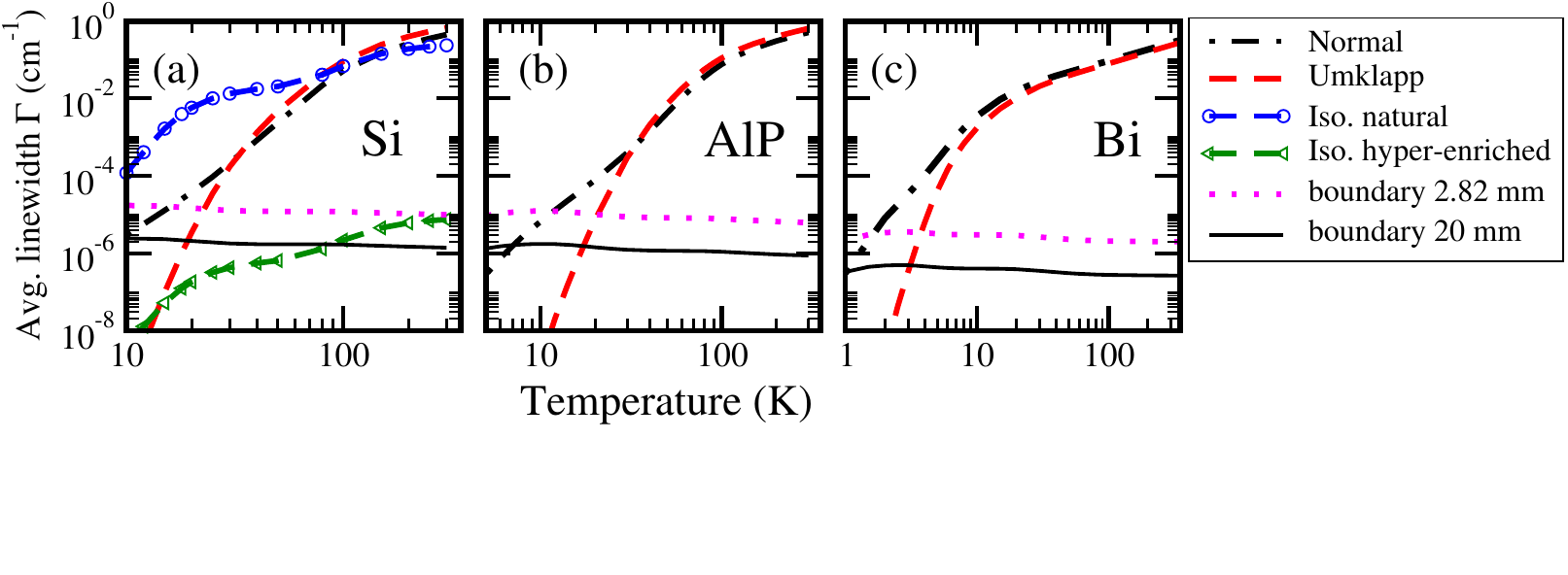}
 \caption{The thermodynamic average of phonon linewith. Black dot-dashed lines - normal processes, red dashed line - Umklapp processes; 
blue and greenlines with symbols - isotopic scattering for natural and hyper-enriched silicon. Horizontal
lines: boundary scattering for 2.82~mm (red dotted line) and 20~mm (black solid line) samples.}
\label{linewidth_Si}
\end{figure*}

To further understand why the effect of repopulation reduction by  boundary scattering is stronger
in Si and AlP compared to Bi, we study the average scattering rates in  Fig.~\ref{linewidth_Si}.
By comparing the Umklapp and normal scattering rates in panels a (Si), b (AlP) and c (Bi) of Fig.~\ref{linewidth_Si},
we notice that overall, in all three materials the normal scattering dominates over the Umklapp scattering at low temperatures.
In that respect, cubic materials studied in this work are similar to Bi. The major role of isotopic scattering, which is the 
dominant scattering process below 70~K in natural silicon, is also illustrated in panel~a of Fig.~\ref{linewidth_Si}.

Coming now to boundary scattering rates, we notice that for equal sample sizes, boundary scattering rates are 4 to 5 times larger
in Si and AlP, compared to Bi. This fact, which is due  to larger phonon group velocities in Si and AlP as compared to those 
in Bi, explains why the repopulation reduction by  boundary scattering is stronger
in Si and AlP.

\section{\label{sec:concl} Conclusions}

In  this work we have performed the theoretical analysis of the
conditions necessary to reach the collective heat transport limit in silicon with
various isotopic compositions, as well as in  AlAs and AlP which contain naturally one single stable isotope.  
While the collective heat transport is impossible in natural silicon
due to  isotopic scattering, it can in principle be reached in hyper-enriched Si, as well as in natural AlAs and AlP.
We have shown that, contrarily to common belief, the normal phonon-phonon scattering processes
play an important role in cubic semiconductors below  20~K, similarly to Bi where collective heat transport phenomena
have been experimentally observed. 
In addition, in the cubic materials studied in this work, 
the possibility to reach the collective heat transport limit is reduced by the large phonon group velocities that enhance 
the effect of boundary scattering with respect to, e.g., bismuth. 
Nevertheless, as we have shown in the present work, the observation of the
collective heat transport limit in AlP must be possible at temperatures around 16 K for 20~mm samples.

\section{Acknowledgements}
Calculations have been performed with the Quantum ESPRESSO computational package~\cite{Giannozzi:2017} and the D3Q code~\cite{Paulatto:2013,Fugallo:2013}.
We are grateful to A.V.~Inyushkin for providing the experimental data published in Refs.~\cite{Inyushkin:2004,Inyushkin:2018}.
We asknowlege useful discussions with B. Fauquet and F. Mauri.
This work has been granted access to HPC resources by he French HPC centers GENCI-IDRIS, GENCI-CINES and GENCI-TGCC (Project 2210) and by the Ecole Polytechnique through the \textit{Three Lab Computing} cluster.
Financial support from the ANR PLACHO project ANR-21-CE50-0008 is gratefully acknowleged. \\

\bibliography{mybiblio}

\begin{thebibliography}{10}

\bibitem{Ding:2022}
Z.~Ding, K.~Chen, B.~Song, J.~Shin, A.~Maznev, K.~Nelson, and G.~Chen.
\newblock Observation of second sound in graphite over 200 {K}.
\newblock {\em Nat. Comm.}, 13:285, 2022.

\bibitem{Sendra:2022}
L.~Sendra, A.~Beardo, J.~Bafaluy, P.~Torres, F.~X. Alvarez, and J.~Camacho.
\newblock Hydrodynamic heat transport in dielectric crystals in the collective
  limit and the drifting/driftless velocity conundrum.
\newblock {\em Phys. Rev. B}, 106:155301, 2022.

\bibitem{Beardo:2021}
A.~Beardo, M.~L\'opez-Su\'arez, L.~A. P\'erez, L.~Sendraand M.~I. Alonso,
  C.~Melis, J.~Bafaluy, J.~Camacho, L.~Colombo, R.~Rurali, F.~X. Alvarez, and
  J.~S. Reparaz.
\newblock Observation of second sound in a rapidly varying temperature field in
  {Ge}.
\newblock {\em Sci. Adv.}, 7:eabg4677, 2021.

\bibitem{Ghosh:2022}
K.~Ghosh, A.~Kusiak, and J.-L. Battaglia.
\newblock Phonon hydrodynamics in crystalline materials.
\newblock {\em J.~Phys.: Condens. Matter}, 34:323001, 2022.

\bibitem{Ghosh:2021}
K.~Ghosh, A.~Kusiak, and J.-L. Battaglia.
\newblock Effect of characteristic size on the collective phonon transport in
  crystalline {GeTe}.
\newblock {\em Phys. Rev. Materials}, 5:073605, 2021.

\bibitem{Beardo:2019}
A.~Beardo, M.~Calvo-Schwarzwalder, J.~Camacho, T.~G. Myers, P.~Torres,
  L.~Sendra, F.~X. Alvarez, and J.~Bafaluy.
\newblock Hydrodynamic heat transport in compact and holey silicon thin films.
\newblock {\em Phys. Rev. App.}, 11:034003, 2019.

\bibitem{Ding:2018}
Z.~Ding, J.~Zhou, B.~Song, V.~Chiloyan, M.~Li, T.-H. Liu, and G.~Chen.
\newblock Phonon hydrodynamic heat conduction and knudsen minimum in graphite.
\newblock {\em Nano Lett.}, 18:638, 2018.

\bibitem{Machida:2020}
Y.~Machida1, N.~Matsumoto, T.~Isono, and K.~Behnia.
\newblock Phonon hydrodynamics and ultrahigh–room-temperature thermal
  conductivity in thin graphite.
\newblock {\em Science}, 367:309, 2020.

\bibitem{Machida:2018}
Y.~Machida, A.~Subedi, K.~Akiba, A.~Miyake, M.~Tokunaga, Y.~Akahama, K.~Izawa,
  and K.~Behnia.
\newblock Observation of {P}oiseuille flow of phonons in black phosphorus.
\newblock {\em Science Advances}, 4:eaat3374, 2018.

\bibitem{Cahill:2015}
D.~G. Cahill, P.~V. Braun, G.~Chen, D.~R. Clarke, S.~H Fan, K.~E. Goodson,
  P.~Keblinski, W.~P. King, G.~D. Mahan, A.~Majumdar, H.~J. Maris, S.~R.
  Phillpot, E.~Pop, and L.~Shi.
\newblock Nanoscale thermal transport. {II}. 2003-2012.
\newblock {\em Applied Physics Reviews}, 1:011305, 2015.

\bibitem{Volz:2016}
S.~Volz, J.~Ordonez-Miranda, A.~Shchepetov, M.~Prunnila, J.~Ahopelto,
  T.~Pezeril, G.~Vaudel, V.~Gusev, P.~Ruello, E.~Weig, M.~Schubert, M.~Hettich,
  M.~Grossman, T.~Dekorsy, F.~Alzina, B.~Graczykowski, E.~Chavez-Angel, J-S.
  Reparaz, M.~R. Wagner, C.~M. Sotomayor-Torres, S.~Xiong, S.~Neogi, and
  D.~Donadio.
\newblock Nanophononics: state of the art and perspectives.
\newblock {\em Eur. Phys. J. B}, 89:15, 2016.

\bibitem{Chang:2008}
C.~W. Chang, D.Cohen Okawa, H.~Garcia, A.~Majumdar, and A.~Zettl.
\newblock Breakdown of {F}ourier's law in nanotube thermal conductors.
\newblock {\em Phys. Rev. Lett.}, 101:075903, 2008.

\bibitem{Yang:2010}
N.~Yand, G.~Zhang, and B.~Li.
\newblock Violation of {F}ourier's law and anomalous heat diffusion in silicon
  nanowires.
\newblock {\em Nano~Today}, 5:85, 2010.

\bibitem{Cepellotti:2016}
A.~Cepellotti and N.~Marzari.
\newblock Thermal transport in crystals as a kinetic theory of relaxons.
\newblock {\em Phys. Rev. X}, 6:041013, 2016.

\bibitem{Cepellotti:2017}
A.~Cepellotti and N.~Marzari.
\newblock Transport waves as crystal excitations.
\newblock {\em Phys. Rev. Mat.}, 1:045406, 2017.

\bibitem{Li:2019}
X.~Li and S.~Lee.
\newblock Crossover of ballistic, hydrodynamic, and diffusive phonon transport
  in suspended graphene.
\newblock {\em Phys. Rev. B}, 99:085202, 2019.

\bibitem{Majee:2018}
A.~K. Majee and Z.~Aksamija.
\newblock Dynamical thermal conductivity of suspended graphene ribbons in the
  hydrodynamic regime.
\newblock {\em Phys. Rev. B}, 98:024303, 2018.

\bibitem{Zhang:2011d}
J.~Zhang, X.~Huang, Y.~Yue, J.~Wang, and X.~Wang.
\newblock Dynamical response of graphene to thermal impulse.
\newblock {\em Phys. Rev. B}, 84:235416, 2011.

\bibitem{Fugallo:2014}
G.~Fugallo, A.~Cepellotti, L.~Paulatto, M.~Lazzeri, N.~Marzari, and F.~Mauri.
\newblock Thermal conductivity of graphene and graphite: collective excitations
  and mean free paths.
\newblock {\em Nano Lett.}, 14:6109, 2014.

\bibitem{Lee:2015}
S.~Lee, D.~Broido, K.~Esfarjani, and G.~Chen.
\newblock Hydrodynamic phonon transport in suspended graphene.
\newblock {\em Nature Communications}, 6:6290, 2015.

\bibitem{Cepellotti:2015}
A.~Cepellotti, G.~Fugallo, L.~Paulatto, M.~Lazzeri, F.~Mauri, and N.~Marzari.
\newblock Phonon hydrodynamics in two-dimensional materials.
\newblock {\em Nature Communications}, 6:6400, 2015.

\bibitem{Huang:2023}
X.~Huang, Y.~Guo, Y.~Wu, S.~Masubuchi, K.~Watanabe, T.~Taniguchi, Z.~Zhang,
  S.~Volz, T.~Machida, and M.~Nomura.
\newblock Observation of phonon {P}oiseuille ﬂow in isotopically puriﬁed
  graphite ribbons.
\newblock {\em Nat. Comm.}, 14:2044, 2023.

\bibitem{Huberman:2019}
S.~Huberman, R.A. Duncan, K.~Chen, B.~Song, V.~Chiloyan, Z.~Ding, A.A. Maznev,
  G.~Chen, and K.A. Nelson.
\newblock Observation of second sound in graphite at temperatures above 100
  {K}.
\newblock {\em Science}, 10.1112:science.aav3548, 2019.

\bibitem{Martelli:2018}
V.~Martelli, J.L. Jimenez, M.~Continentino, E.~Baggio-Saitovitch, and
  K.~Behnia.
\newblock Thermal transport and phonon hydrodynamics in strontium titanate.
\newblock {\em Phys. Rev. Lett.}, 120:125901, 2018.

\bibitem{Rabe:2007}
K.~M. Rabe, C.~H. Ahn, and J.-M. Triscone.
\newblock {\em Physics of Ferroelectrics. A Modern Perspective}.
\newblock Springer-Verlag, Berlin Heidelberg, 2007.

\bibitem{Gurevich:1988}
V.~L. Gurevich and A.~K. Tagantsev.
\newblock Second sound in ferroelectrics.
\newblock {\em Sov. Phys. JETP}, 67:206--212, 1988.

\bibitem{Tiberkevich:2019}
V.~Tiberkevich, I.~V. Borisenko, P.~Nowik-Boltyk, V.~E. Demidov, A.~B.
  Rinkevich, S.~O. Demokritov, and A.~N. Slavin.
\newblock Excitation of coherent second sound waves in a dense magnon gas.
\newblock {\em Scientific Reports}, 9:9063, 2019.

\bibitem{Narayanamurti:1972}
V.~Narayanamurti and R.C. Dynes.
\newblock Observation of second sound in bismuth.
\newblock {\em Phys. Rev. Lett.}, 28:1461, 1972.

\bibitem{Mezhov:1974}
L.P. Mezhov-Deglin, V.N. Kopylov, and E.S. Medvedev.
\newblock Contributions of various phonon relaxation mechanisms to the thermal
  resistance of the crystal lattice of bismuth at temperatures below 2~{K}.
\newblock {\em Sov. Phys. JETP}, 40:557, 1974.

\bibitem{Ackerman:1966}
C.~C. Ackerman, B.~Bertman, H.~A. Fairbank, and R.~A. Guyer.
\newblock Second sound in solid helium.
\newblock {\em Phys. Rev. Lett.}, 16:789, 1966.

\bibitem{Jackson:1970}
H.~E. Jackson, C.~T. Walker, and T.~F. McNelly.
\newblock Second sound in {NaF}.
\newblock {\em Phys. Rev. Lett.}, 25:26, 1970.

\bibitem{Pohl:1976}
D.~W. Pohl and V.~Irniger.
\newblock Observation of {S}econd {S}ound in {NaF} by {M}eans of {L}ight
  {S}cattering.
\newblock {\em Phys. Rev. Lett.}, 36:480, 1976.

\bibitem{Thacher:1967}
P.~D. Thacher.
\newblock Effect of boundaries and isotopes on thermal conductivity of {LiF}.
\newblock {\em Phys. Rev.}, 156:975, 1967.

\bibitem{Ozhogin:1996}
V.~I. Ozhogin, A.~V. Inyushkin, A.~N. Taldenkov, A.~V. Tikhomirov, and G.~E.
  Popov.
\newblock Isotope effect in the thermal conductivity of germanium single
  crystals.
\newblock {\em JETP Lett.}, 63:490, 1996.

\bibitem{Zhernov:2002}
A.~P. Zhernov and A.~V. Inyushkin.
\newblock Kinetic coefficients in isotopically disordered crystals.
\newblock {\em Phys. Usp.}, 45:527--555, 2002.

\bibitem{Lindsay:2016}
L.~Lindsay.
\newblock First principles {P}eierls-{B}oltzmann phonon thermal transport: A
  topical review.
\newblock {\em Nanoscale and Microscale Thermophysical Engineering}, 20:67,
  2016.

\bibitem{Inyushkin:2004}
A.~V. Inyushkin, A.~N. Taldenkov, A.~M. Gibin, A.~V. Gusev, and H.-J. Pohl.
\newblock On the isotope effect in thermal conductivity of silicon.
\newblock {\em Phys. Status Solidi C}, 1:2995, 2004.

\bibitem{Inyushkin:2018}
A.~V. Inyushkin, N.~V. Abrosimov, A.~N. Taldenkov, J.~W. Ager, E.~E. Haller,
  H.~Riemann, H-J. Pohl, and P.~Becker.
\newblock Ultrahigh thermal conductivity of isotopically enriched silicon.
\newblock {\em J. Appl. Phys.}, 123:095112, 2018.

\bibitem{Markov:2018}
M.~Markov, J.~Sjakste, G.~Barbarino, G.~Fugallo, L.~Paulatto, M.~Lazzeri,
  F.~Mauri, and N.~Vast.
\newblock Hydrodynamic heat transport regime in bismuth : a theoretical
  viewpoint.
\newblock {\em Phys. Rev. Lett.}, 120:075901, 2018.

\bibitem{Abrosimov:2017}
N.V. Abrosimov, D.G. Aref'ev, P.~Becker, H.~Bettin, A.D. Bulanov, M.~F.
  Churbanov, S.V. Filimonov, V.A. Gavva, O.N. Godisov, A.V. Gusev, T.V.
  Kotereva, D.~Nietzold, M.~Peters, A.M. Potapov, H.-J. Pohl, A.~Pramann,
  H.~Riemann, P.-T. Scheel, R.~Stosch, S.~Wundrack, and S.~Zakel.
\newblock A new generation of 99.999 enriched 28 {S}i single crystals for the
  determination of {A}vogadro's constant.
\newblock {\em Metrologia}, 54:599, 2017.

\bibitem{Fhipp}
M.~Fuchs and \textit{et al.}
\newblock \url{http://www.fhi-berlin.mpg.de/th/fhi98md/fhi98PP/}.

\bibitem{Baroni:2001}
S.~Baroni, S.~de~Gironcoli, A.~Dal Corso, and P.~Giannozzi.
\newblock Phonons and related crystal properties from density-functional
  perturbation theory.
\newblock {\em Rev. Mod. Phys.}, 73:515, 2001.

\bibitem{Giannozzi:2017}
P.~Giannozzi, O.~Andreussi, T.~Brumme, O.~Bunau, M.~Buongiorno Nardelli,
  M.~Calandra, R.~Car, C.~Cavazzoni, D.~Ceresoli, M.~Cococcioni, N.~Colonna,
  I.~Carnimeo, A.~Dal Corso, S.~de~Gironcoli, P.~Delugas, R.~A. DiStasio
  Jr.and~A. Ferretti, A.~Floris, G.~Fratesi, G.~Fugallo, R.~Gebauer,
  U.~Gerstmann, F.~Giustino, T.~Gorni, J.~Jia, M.~Kawamura, H.-Y. Ko,
  A.~Kokalj, E.~K\"u\c{c}\"ukbenli, M.~Lazzeri, M.~Marsili, N.~Marzari,
  F.~Mauri, N.~L. Nguyen, H.-V. Nguyen, A.~Otero de-la Roza, L.~Paulatto,
  S.~Ponc\'e, D.~Rocca, R.~Sabatini, B.~Santra, M.~Schlipf, A.~P. Seitsonen,
  A.~Smogunov, I.~Timrov, T.~Thonhauser, P.~Umari, N.~Vast, X.~Wu, and
  S.~Baroni.
\newblock Advanced capabilities for materials modelling with \textsc{Quantum
  ESPRESSO}.
\newblock {\em J. Phys.: Condens. Matter}, 29:465901, 2017.

\bibitem{Paulatto:2013}
L.~Paulatto, F.~Mauri, and M.~Lazzeri.
\newblock Anharmonic properties from a generalized third-order \textit{ab
  initio} approach: Theory and applications to graphite and graphene.
\newblock {\em Phys. Rev. B}, 87:214303, 2013.

\bibitem{Fugallo:2013}
G.~Fugallo, M.~Lazzeri, L.~Paulatto, and F.~Mauri.
\newblock \textit{Ab initio} variational approach for evaluating lattice
  thermal conductivity.
\newblock {\em Phys. Rev. B}, 88:045430, 2013.

\bibitem{Omini:1995}
M.~Omini and A.~Sparavigna.
\newblock An iterative approach to the phonon {B}oltzmann equation in the
  theory of thermal conductivity.
\newblock {\em Physica B}, 212:101, 1995.

\bibitem{Sparavigna:2002}
A.~Sparavigna.
\newblock Influence of isotope scattering on the thermal conductivity of
  diamond.
\newblock {\em Phys. Rev. B}, 65:064305, 2002.

\bibitem{Markov:2016}
M.~Markov, J.~Sjakste, G.~Fugallo, L.~Paulatto, M.~Lazzeri, F.~Mauri, and
  N.~Vast.
\newblock Nanoscale mechanisms for the reduction of heat transport in bismuth.
\newblock {\em Phys. Rev. B}, 93:064301, 2016.

\bibitem{Sen:2023}
R.~Sen, N.~Vast, and J.~Sjakste.
\newblock Role of dimensionality and size in controlling the drag seebeck
  coefficient of doped silicon nanostructures: A fundamental understanding.
\newblock {\em Phys. Rev. B}, 108:L060301, 2023.

\bibitem{Tamura:1983}
S.~Tamura.
\newblock Isotope scattering of dispersive phonons in {G}e.
\newblock {\em Phys. Re. B}, 27:858, 1983.

\bibitem{Omini:1997}
M.~Omini and A.~Sparavigna.
\newblock Heat transport in dielectric solids with diamond structure.
\newblock {\em Il Nuovo Cimento D}, 19:1537, 1997.

\bibitem{Ward:2009}
A.~Ward, D.~A. Broido, D.~A. Stewart, and G.~Deinzer.
\newblock \textit{Ab initio} theory of the lattice thermal conductivity in
  diamond.
\newblock {\em Phys.\ Rev.~B}, 80:125203, 2009.

\bibitem{Lindsay:2013b}
L.~Lindsay, D.~A. Broido, and T.~L. Reinecke.
\newblock Ab initio thermal transport in compound semiconductors.
\newblock {\em Phys. Rev. B}, 87:165201, 2013.

\end{thebibliography}
\bibliographystyle{unsrt}

\end{document}